# Integrated Real-Time Motion Tracking and AI Analysis for Athletic Performance Optimization

Parth Agrawal
*Department of Computer Science and Engineering, Chandigarh University*
Punjab, India
parthworks247@gmail.com
0009-0003-7983-9107

Ronit
*Department of Computer Science and Engineering, Chandigarh University*
Punjab, India
rronit801@gmail.com

Vinod Kumar
*Department of Computer Science and Engineering, Chandigarh University*
Punjab, India
dcsavinod@gmail.com

Sagar Kumar
*Department of Computer Science and Engineering, Chandigarh University*
Punjab, India
sagarsaxena514@gmail.com

Aashish Bhambri
*Department of Computer Science and Engineering, Chandigarh University*
Punjab, India
aashishbhambri03@gmail.com

*Abstract*— **Applying Human Pose Estimation (HPE) in real world environments remains a challenging task, this paper explores and surveys real time HPE approaches and their limitations in sports analysis for individuals, alongside developing a practical lightweight prototype for real world testing and usage. The older marker-based motion capture systems evolving to the modern accessible and adaptable markerless deep learning approaches, this survey explores the foundational architectures, which balance precision and efficiency. We also compare algorithmic frameworks (top-down, bottom-up, one-stage approaches, etc.) on practical deployment metrics such as inference latency, frame rate, mean per-joint position error, and temporal jitter to guide model selection process for sports application. As our prime contribution, we are proposing a modular, lightweight software prototype, which uses MediaPipe HPE framework with multiple exercise specific logic to deliver real-time insights and AI based feedback for non-expert users. We derive sports insights and providing feedback with minimal computational resources, while showcasing the performance and reliability metrics. In the end, we suggest other future research directions like combining sensors, and AR/VR. This work caters to researchers, engineers, sport scientists, etc., as both technical resource and a valid blueprint to implement a similar or improved real-time HPE analysis system for athletic performance enhancement or other purposes.**

## I. INTRODUCTION

Coaches, scouts, etc. have always used intuition and manual observation to analyse athlete performance. This subjective and labour-intensive method often missed the subtle, high-speed nuances as manually reviewing footage, logging, and observing information was very time-consuming [1]. The field of Human Pose Estimation (HPE) focuses on identifying and tracking the anatomical key points of the human body [2]. Early sports science research studies used marker-based optoelectronic systems on athletes to achieve sub-millimetric accuracy [8], but its practical use was limited due to expensive, controlled, lengthy and intrusive lab setups, which may disrupt natural body movement, i.e., largely unusable or useless for proper in-game analysis, for a great majority of teams and athletes [8].

The development of Marker-less HPE and evolution in motion capture technology and computational power eliminated the need for physical markers by leveraging deep learning-based algorithms to localize human anatomical key points, allowing us to capture and process player data from videos [8] with unprecedented scale and precision, revolutionizing sports performance analytics insight systems [2], while being affordable, non-intrusive, portable, and accessible via common hardware like webcams and phone cameras [8]. Advanced analysis tools available only to well-funded research institutions and professional organizations became accessible to small clubs, and even common individuals like amateurs, athletes and enthusiasts [12].

While offline analysis after an event offers valuable insights, the main advantage of modern HPE lies in its real-time capabilities. Estimating the pose from raw data with ultra-low latency enabled real time data centric insight generation, allowing near instant feedback loop for athletes and coaches during trainings, live events, etc., allowing them to adjust their form [2]. These systems can also help in detecting abnormal movement patterns earlier, flagging fatigue and injury risk, enabling timely intervention to protect athlete [2]. This motivated us to think, *how well can existing frameworks perform in sports performance analysis in real world in providing intelligent, actionable feedback?*

To find the answer of this question, we are proposing a setup, demonstrating its potential in real world. The contributions of this paper are twofold. (1) The study presents a comprehensive overview of the current technical landscape, identify challenges, and outline future research direction. (2) Propose an efficient practical prototype for analysis of athlete's performance, specifically focusing on designing a reusable modular architecture and evaluating its real-time viability in practical settings. and provide intelligent effective feedback to them. Unlike existing HPE-based fitness systems primarily focusing on pose visualization or providing basic data, our proposed system unifies modular low compute exercise analysers, finite state machine-based repetition validation, peak joint angle tracking, and AI-driven post-session feedback for non-expert users.

## II. RELATED WORK

The field of HPE via vision extends from classic model-based methods to dominant deep-learning approaches. Earlier work used to rely on handcrafted features and explicit body models [10]. In these frameworks, the human body is visualized as a complex object composed of rigid parts connected by joints [10]. Despite significant advancements in HPE, existing works either prioritize high accuracy at the cost of computational complexity or focus on offline analysis. Lightweight real time systems with modular and extensible architectures for exercise-specific feedback remain relatively underexplored. This gap motivates the design of our proposed system.

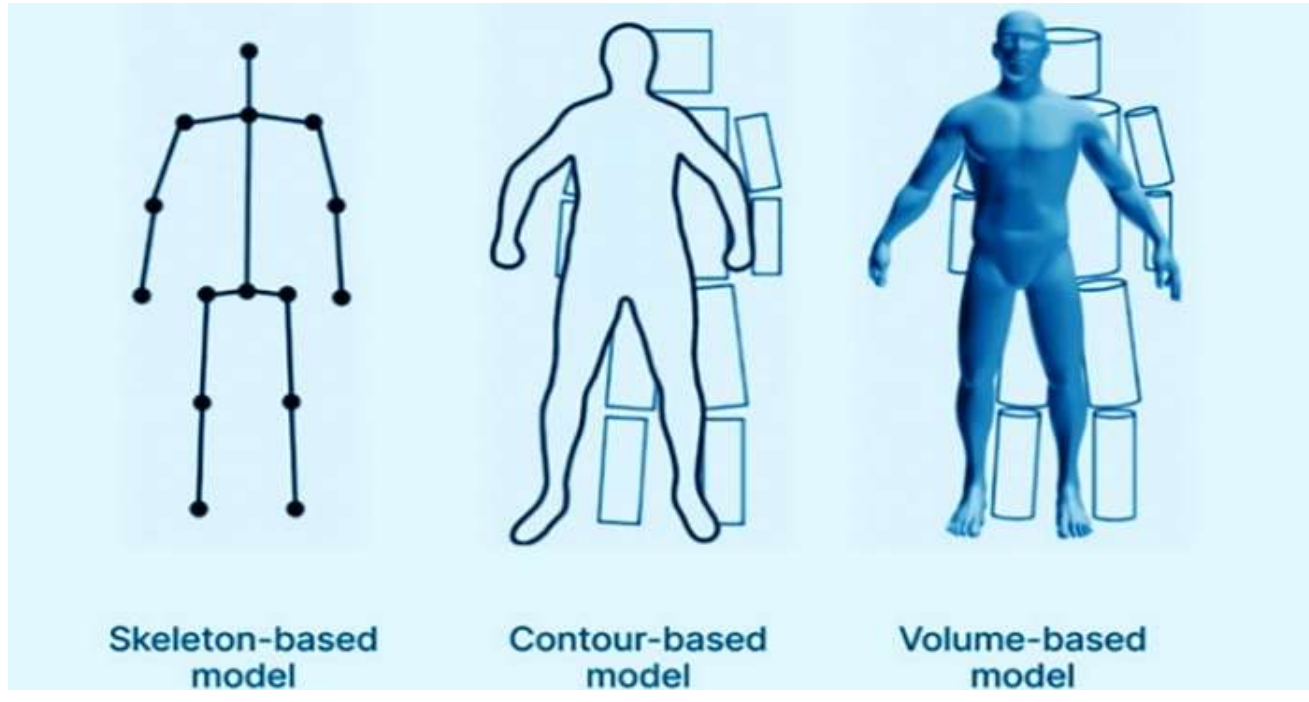

Fig. 1. Types of Human Pose Estimation (HPE) Models [7]

Table 1: Comparison of SOTA human pose estimation approaches

| Approach (Core Idea) | Key Models | Advantages/Role | Limitations and Challenges |
|---|---|---|---|
| **Classic Model-Based Methods** (Handcrafted features with explicit body models [10]) | • Kinematic Models (Pictorial Structures)<br>• Contour-based Models (Active Shape Models)<br>• Volumetric Models (cylinders/cones) | • Approach-wide: Provided basic frameworks for pose estimation<br>• Kinematic Models: Representing body as joints and limbs [5,10]<br>• Contour-based Models: Represent body shape with outlines [5]<br>• Volumetric Models: Introduced the basis for 3D reasoning. | • Kinematic Models: Cannot incorporate texture/shape information effectively [5]<br>• Contour-based Models: Struggle with high degrees of freedom and background clutter [5]<br>• Volumetric Models: Computationally intensive for their time. |
| **Deep Learning and Single-Person Pose Estimation** (End-to-end feature learning from pixels [1]) | • DeepPose [3,9]<br>• Stacked Hourglass<br>• Convolutional Pose Machines | • Approach-wide: CNNs removed need for hand-crafted features [1].<br>• DeepPose: Proved end to end learning to be viable for HPE tasks through iterative refinement [9]<br>• Heatmap Methods (like Hourglass): Provides better spatial generalization and accuracy than direct regression. | • Approach-wide: Early methods focused on single-person, necessitating new strategies for multi-person scenes [1]<br>• Heatmap Methods: Can be computationally expensive due to high-resolution feature maps. |
| **Multi-Person 2D Pose Estimation** (Detect-then-estimate; detect-then-group; simultaneous detection) | • Top-Down: YOLO, Faster R-CNN, CPN, HRNet [1]<br>• Bottom-Up: OpenPose (PAFs) [6], Associative Embedding<br>• One-Stage: SPGR | • Top-Down: Modular, and leverages the best detectors and estimators [1]<br>• Bottom-Up: Remains robust to crowd size, enabling real-time performance [6]<br>• One-Stage: Balances speed, and accuracy by avoiding individual detection, grouping stages together. | • Top-Down: Suffers from early commitment problem; cost scales with people [6,19]<br>• Bottom-Up: Association step becomes difficult with overlapping people [19]<br>• One-Stage: Usually less accurate than top-down methods for complex scenes. |
| **Transformer-Based Architectures** (Capture long-range dependencies with ViTs) | • TokenPose<br>• TransPose<br>• HRFormer | • Approach-wide: Great to model a global relationship between key points to capture implicit body structure.<br>• HRFormer: Integrates the global context with a high-resolution representation. | • Approach-wide: Requires large training data; less effective on small datasets without pretraining.<br>• TransPose: Pure transformer architecture can have higher computational complexity than CNNs. |
| **3D Human Pose Estimation** (Lifting from 2D; multi-view triangulation; mesh recovery; direct regression) | • FCN (Martinez et al.) [8]<br>• LSTM (Hossain and Little)<br>• TCN (Pavllo et al.) [8]<br>• Bone modules (Chen et al.)<br>• GraphCNNs<br>• SMPL/SMPL-X [4] | • 2D-to-3D Lifting: Decouples tasks to simplify the inference [8].<br>• GraphCNNs: Model skeletal constraints naturally.<br>• Multi-View: Resolves depth ambiguity for accurate data [8, 11]<br>• HMR (SMPL): Provides full body shape and surface geometry [4] | • 2D-to-3D Lifting: Ill-posed problem due to depth ambiguity [10].<br>• Temporal methods (LSTM/TCN) can limit connectivity [8]<br>• Multi-View: Requires specialized hardware and calibration [11]<br>• HMR: Parameter optimization is complex; struggles with occlusions.<br>• Direct Regression: Highly ill-posed; needs large 3D training datasets. |
| **One-Stage Multi-Person 3D Mesh Recovery** (Parallel mesh estimation and person detection) | • InstaHMR<br>• ROMP | • InstaHMR: Instance-aware features with attention preserve person-specific information, improving accuracy for overlapping individuals.<br>• Approach-wide: More efficient than multi-stage pipelines. | • Approach-wide: Complex network design with multiple loss functions.<br>• General: Balancing detection and mesh recovery tasks is challenging; still a new area with fewer mature implementations. |
| **Lightweight and Efficient Models** (Compressed models for edge deployment) | • MobilePose<br>• Lite-HRNet<br>• Knowledge distillation frameworks | • Approach-wide: Enable real-time performance on mobile/embedded devices.<br>• Lite-HRNet: Maintains reasonable accuracy with significantly fewer parameters. | • Approach-wide: Accuracy trade-off compared to full-scale models.<br>• Knowledge Distillation: Requires pretrained teacher models.<br>• General: May struggle with complex poses or occluded scenarios. |
| **Emerging Approaches** (Cutting-edge techniques) | • ScoreHMR, DiffPose (Diffusion Models)<br>• NeRF<br>• Self-Supervised Learning<br>• Contrastive Learning methods | • ScoreHMR: Strong performance on ill-posed inverse problems; robust to occlusions.<br>• NeRF: Enables photorealistic rendering and 3D understanding from multi-view data.<br>• Self-Supervised: Reduces dependency on labelled data. | • ScoreHMR: Computationally expensive iterative sampling.<br>• NeRF: Requires multi-view or sequential data; slow training/rendering.<br>• Self-Supervised: Still lags behind supervised methods on benchmarks. |

## III. Methodology

The proposed architecture follows a sequential processing framework beginning with video acquisition and ending with AI-driven feedback generation. The system begins with capturing frames from a video source (sample videos [13]) then captures frames using OpenCV and converting them from BGR to RGB and passes them to MediaPipe framework to estimate body landmarks for a person. A pose landmark detection step is then performed. If no human landmarks are detected, further processing is skipped for that frame. When landmarks are successfully detected, the system dynamically selects an exercise analyser, implemented as independent components, allowing modularity. Relevant joint coordinates are extracted depending on the selected exercise, and angles are computed using triplets of key points. The loop continues until termination is triggered, after which the system releases resources and generates feedback. We designed a prompt pipeline that converts biomechanical signals into structured coaching feedback using constraint-based reasoning. In addition to real-time angle computation, the system records peak joint angles (maximum flexion and extension) during each repetition, providing deeper insight into movement quality and range of motion. Unlike traditional systems that prioritize accuracy or computational cost while only providing raw pose data, our system ensures deterministic mapping between physical performance metrics and generated comprehensive, AI-driven feedback, allowing effective debugging and fallback. MediaPipe is used due to its cross-platform compatibility, efficient real-time performance and Qwen 3.5 4B for efficient feedback generation, reducing computational overhead for the analysis, and improving accessibility for non-expert users.

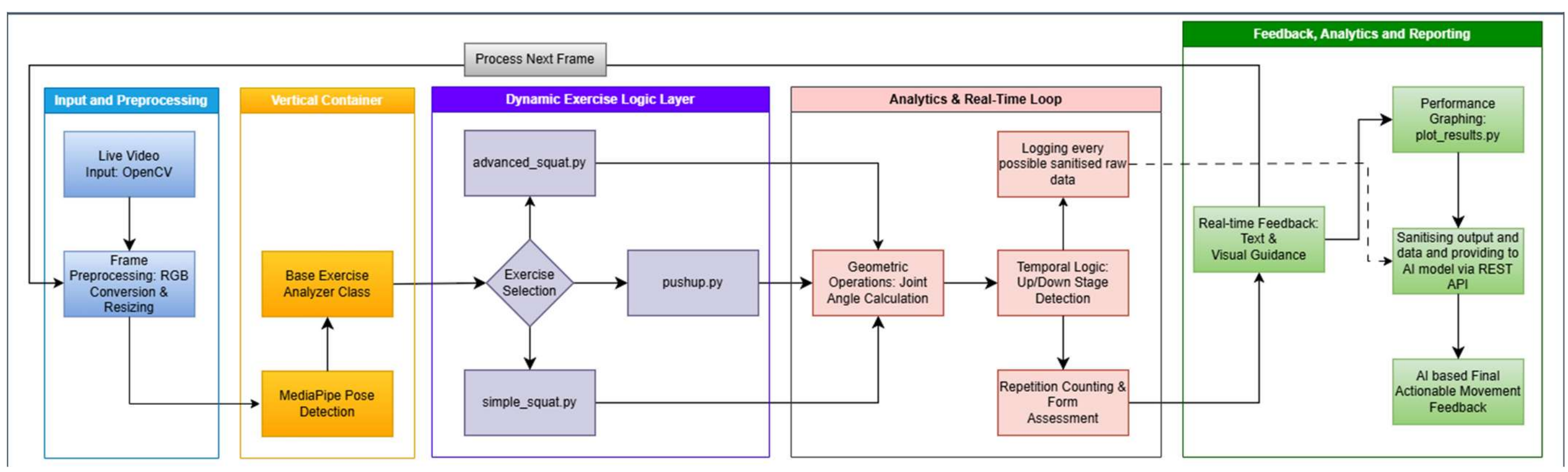

Fig. 2. Diagram for our Prototype for Real-Time Motion Tracking, Analysis and Performance Optimization for Athletes

## IV. Experiments And Results

Table 1: Comparison of current Pose Estimation models, trade-offs between inference time, frame rate, mean per-joint position error and frame jitter.

| Model | Inference Time (ms) | FPS | MPJPE (Relative) | Frame Jitter |
|---|---|---|---|---|
| MediaPipe | 36.29 | 25.7 | 0.0147 | Low |
| Lite-HRNet-30 | ~25–40 | ~25–35 | Slightly higher | Low–Medium |
| Dite-HRNet-30 | ~25–40 | ~25–35 | Comparable | Medium |
| Hourglass | ~150–300 | ~3–7 | Higher | High |
| YOLOv5-Pose | ~40–70 | ~15–25 | Slightly higher | Medium |
| OpenPose | ~120–200 | ~5–10 | Higher | Medium |
| HRNet-W32 | ~50–90 | ~10–20 | Lower | Low |
| HRNet-W48 | ~80–130 | ~8–15 | Lower | Low |
| HigherHRNet-W48 | ~120–200 | ~5–10 | Lower | Low |
| TokenPose-L | ~70–110 | ~10–15 | Lower | Low |
| TransPose-H-A6 | ~70–100 | ~10–15 | Lower | Low |
| ViTPose-B | ~90–140 | ~7–12 | Lower | Low |
| ResNet-152 | ~80–110 | ~8–12 | Comparable | Medium |
| HRFormer-B | ~90–120 | ~7–12 | Lower | Low |

As shown in Table 1, replacing our MediaPipe based system with other models would offer different trade-offs in performance and compatibility. The major advantage of MediaPipe is significantly lower computational cost with cross-platform compatibility, making it suitable for edge deployment. In Fig. 3a and 3b below, we show a snapshot of our prototype in action. It shows completed repetitions, peak high and low metrics, simple feedback and the main parameters for that exercise (Hindu Squat here). Fig. 3a shows him going down while Fig. 3b shows him going up.

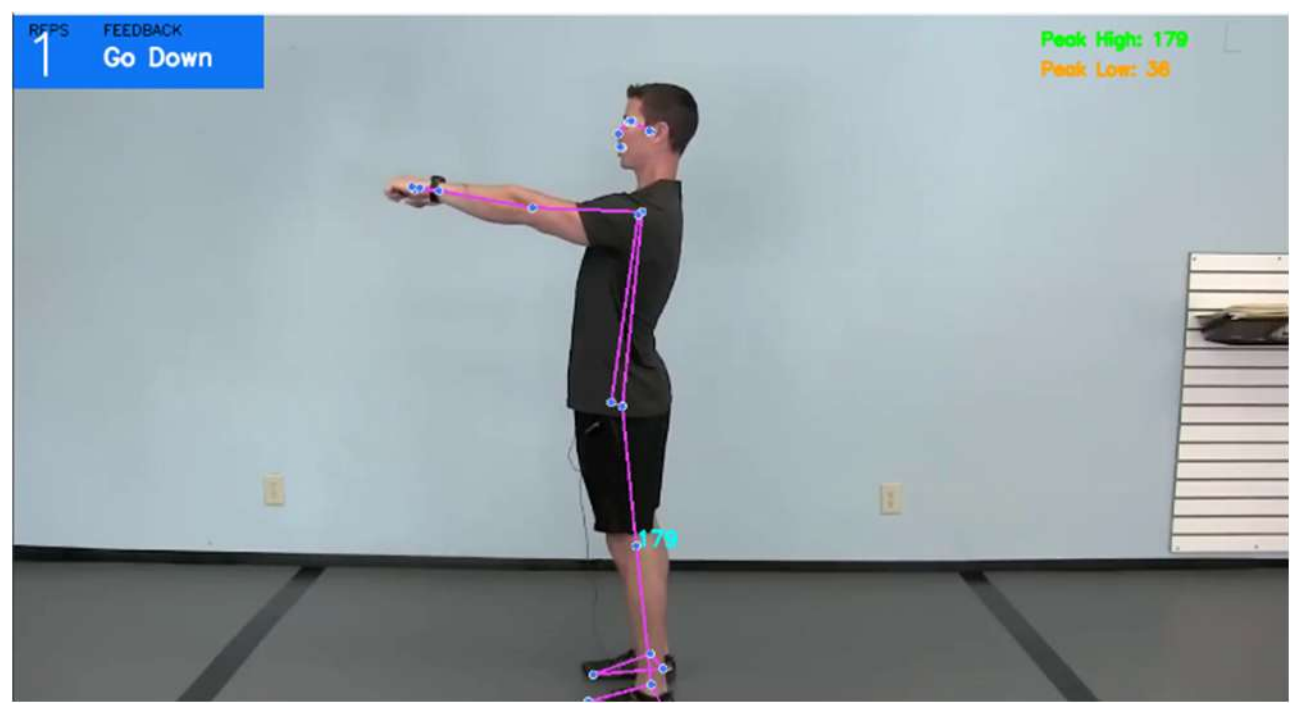

Fig. 3a. Performing Hindu Squat (Going Down from rest)

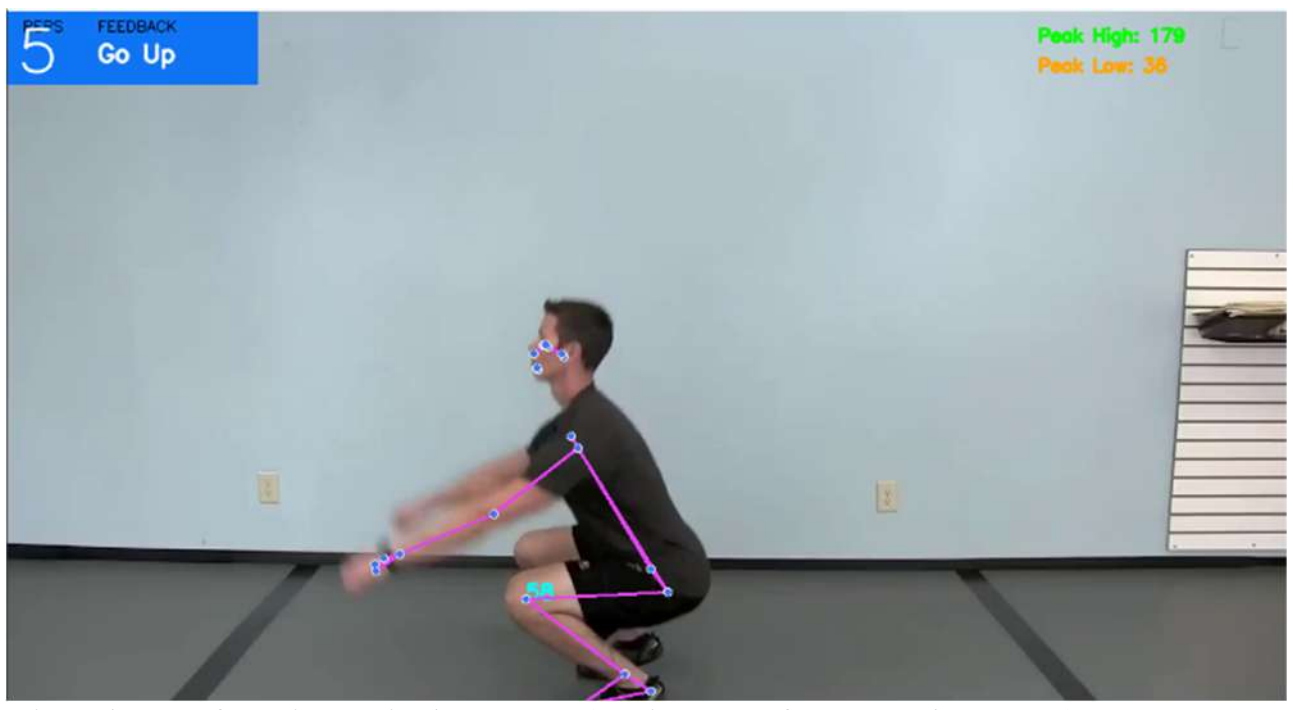

Fig. 3b. Performing Hindu Squat (Going Up from Peak Down)

We evaluated proposed prototype [14] on four different sample videos [13] to assess its performance across multiple conditions, including variations in motion dynamics, camera angles, and execution styles. The evaluation focuses on key system-level metrics such as pose estimation accuracy, processing latency, stability, and repetition counting reliability. Hardware used is a laptop (RTX 3060 + 5800H).

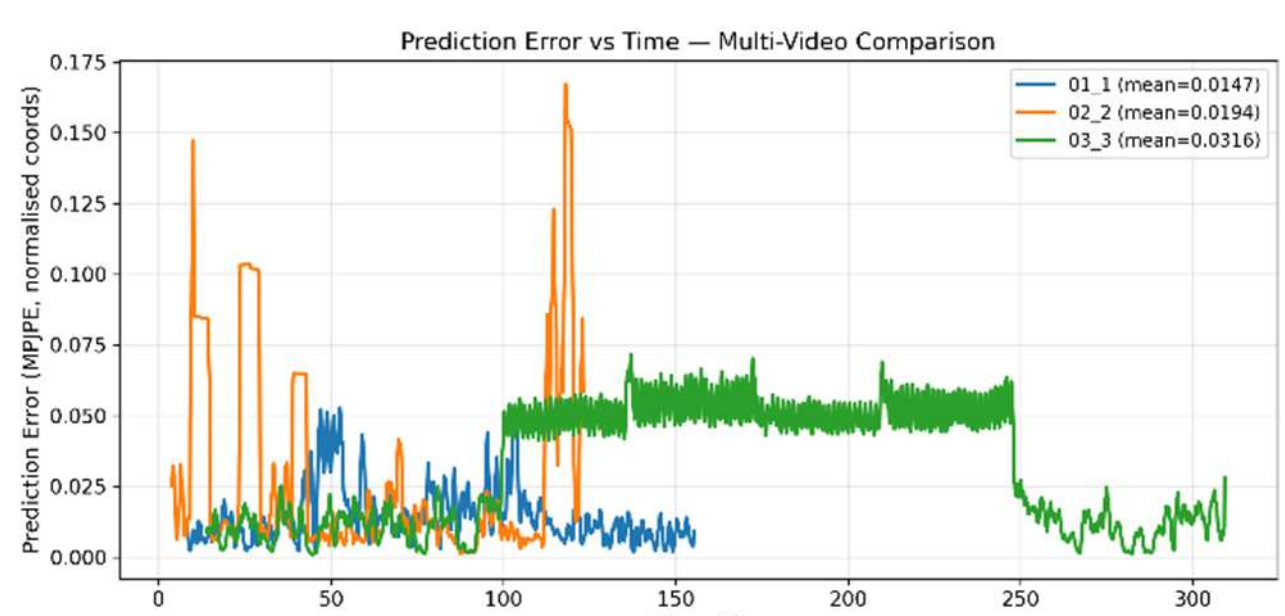

Fig. 4. Prediction Error Timeseries

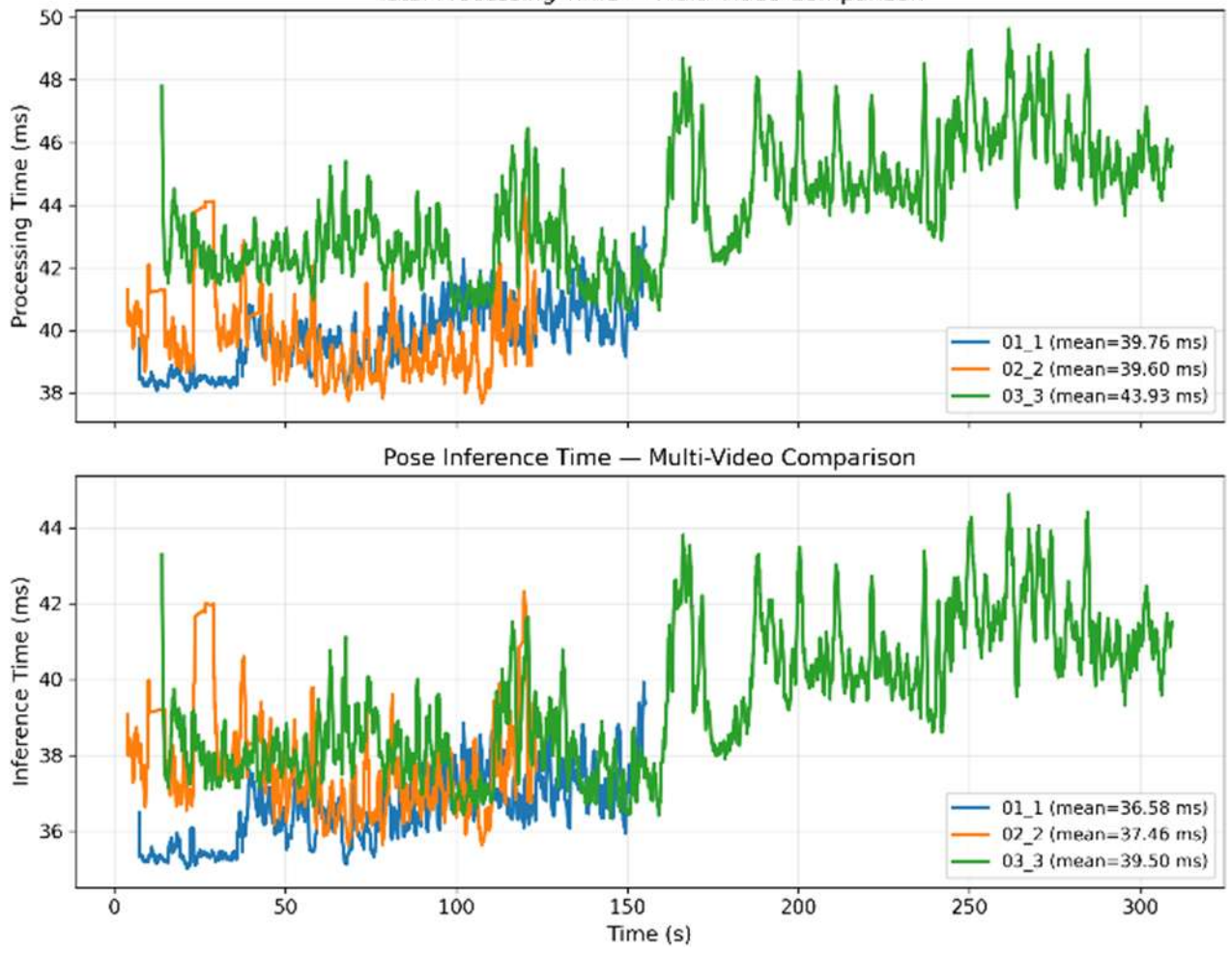

Fig. 5. Processing Time

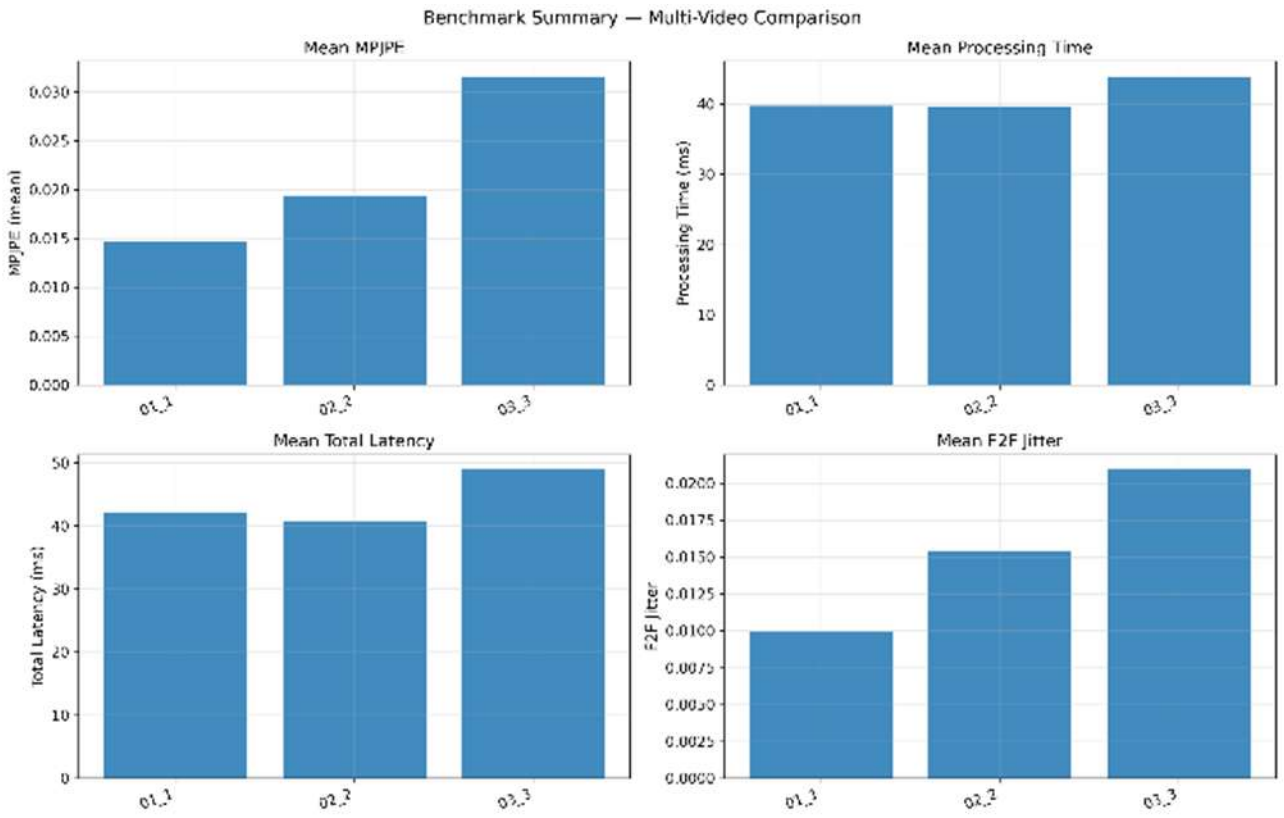

Fig. 6. Mean Per Joint Position Error (MPJPE) per Joint – 33 Key points

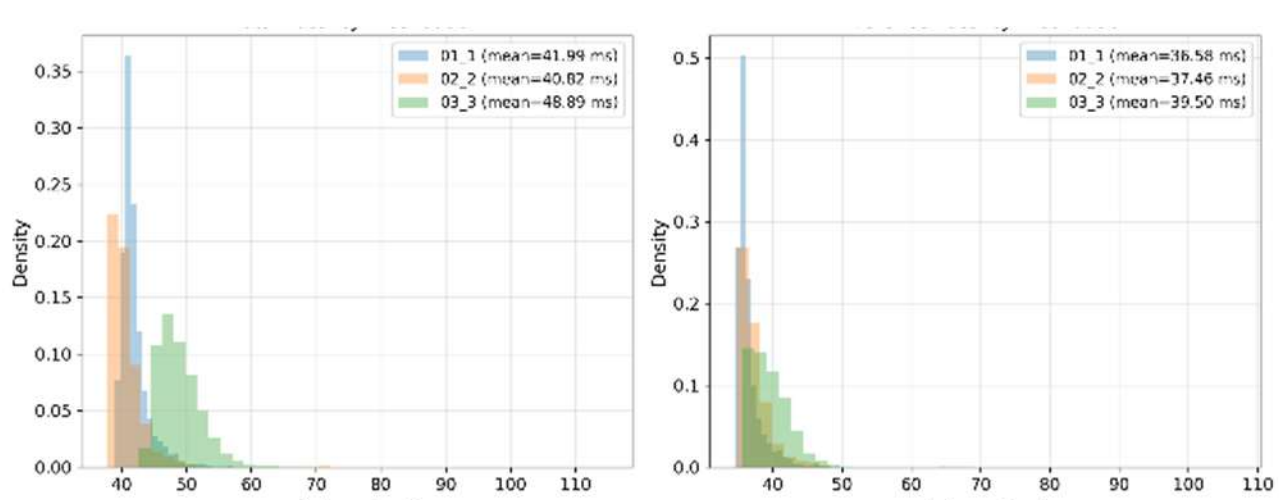
Fig. 7. Total Latency (Left) and Pose Inference Latency (Right)

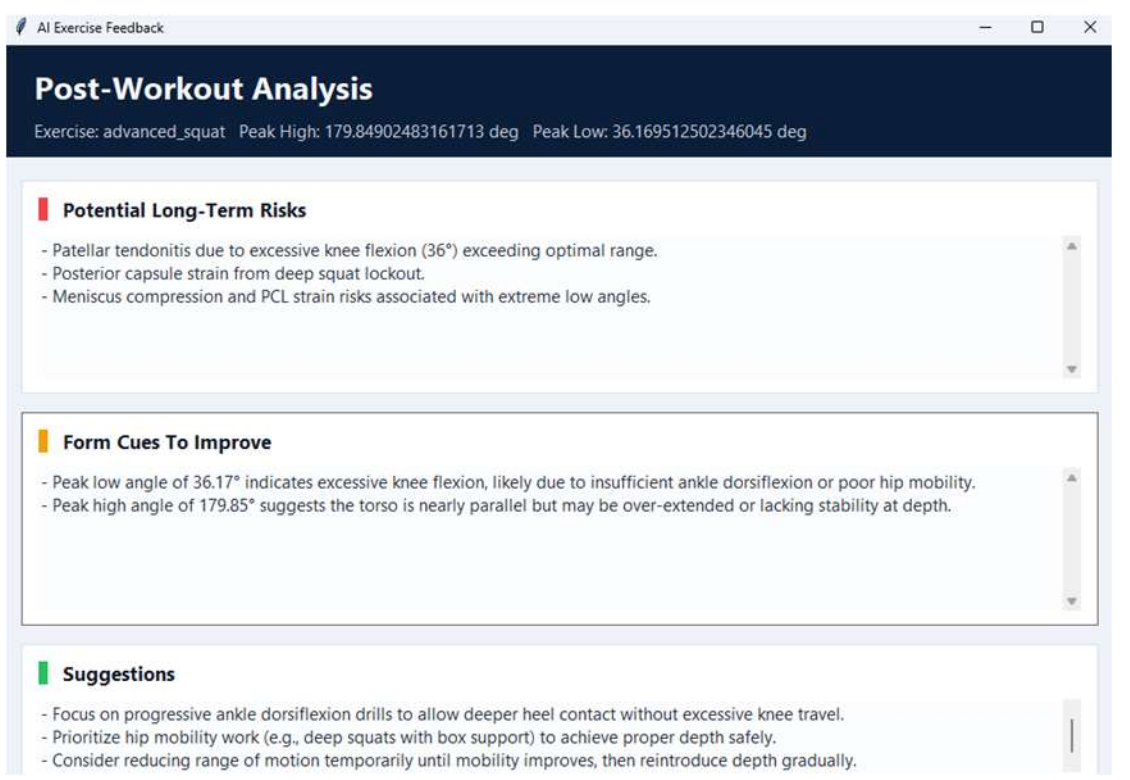

Fig. 8. Post Workout AI Analysis Feedback (For 1st Sample Video [13])

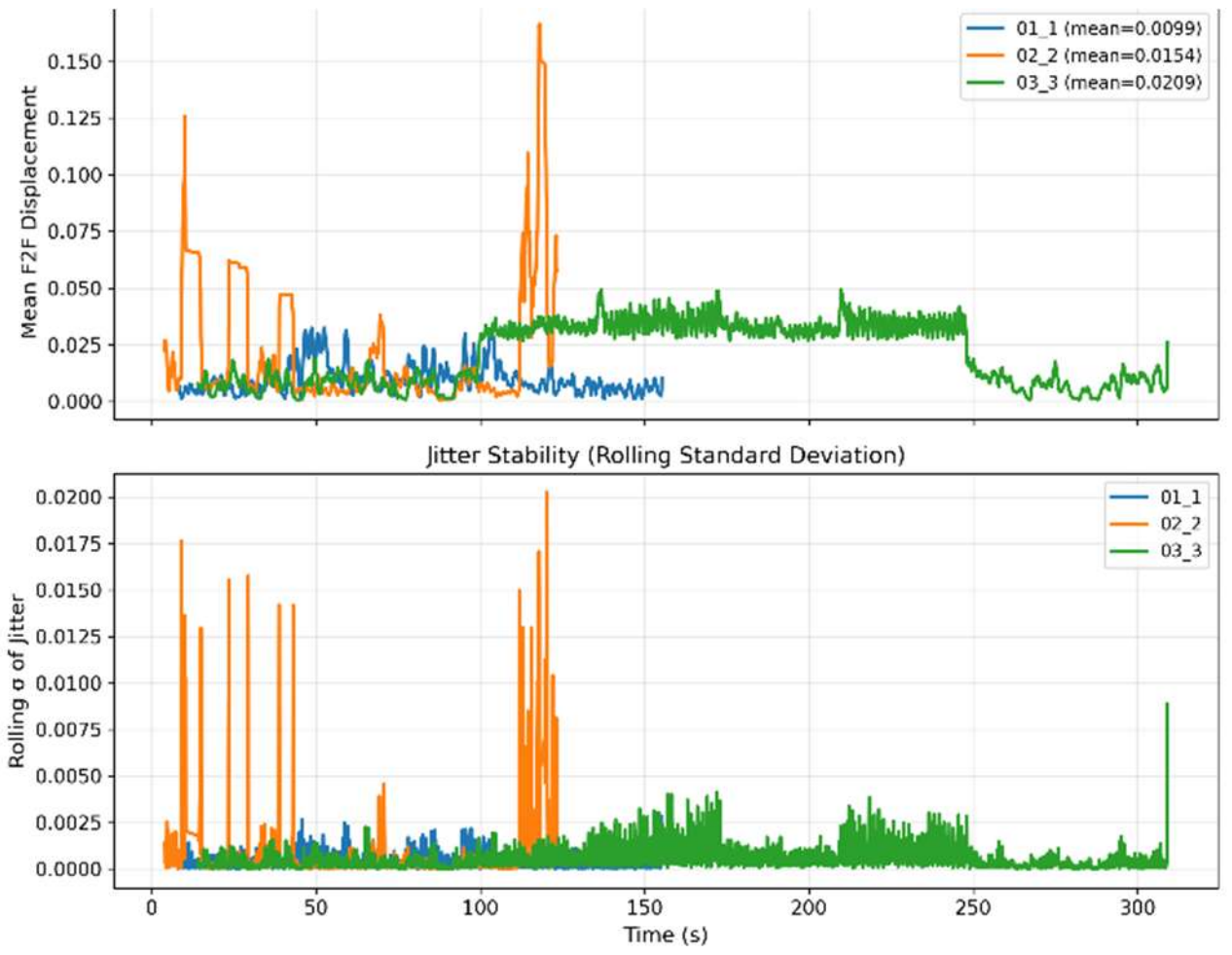

Fig. 9. Frame-to-Frame Jitter / Error Stability

Based on the multi-sample evaluation summarized in Figures 4–9, the prototype demonstrated consistent performance under varied real-world conditions across four recorded exercise sessions with varying lighting, clothing, and movement tempos, the normalized Mean Per Joint Position Error (MPJPE) remained low, ranging from 0.0147 to 0.0316 (Fig. 4 and Fig. 6), confirming stable landmark localization, while the per joint analysis showed that, even though most anatomical key points maintain low error values, extreme parts like fingers and thumbs show higher values, probably due to their smaller spatial representation in the image and increased susceptibility to occlusion and motion blur. The system efficiency is proven to be robust. Mean per-frame processing time across all trials ranged from 39.60ms to 43.93ms (approximately 23–25 FPS), with the MediaPipe inference pass accounting for the majority of this budget (Fig. 7). Total end-to-end latency remained tightly distributed, with 95th-percentile values under 45ms in all cases (Fig. 7), underscoring the pipeline's suitability for real-time interactive feedback. Frame-to-frame jitter, measured via rolling standard deviation of landmark displacement, was consistently low across all sessions (Fig. 9), ensuring that overlaid visual cues and angle readouts appear smooth and stable to the user. After completion of the exercise session, performance data including repetition count, joint angle ranges, etc. are passed to a language model interface where the model generates personalized feedback highlighting form corrections and performance insights (as shown in Fig. 8). Repetition counting remained consistent across all evaluated videos, aligning closely with observed motion cycles and demonstrating reliable temporal segmentation.

## V. Challenges And limitations

In this paper, we present a comprehensive overview of the current technical landscape, identified the current challenges, and proposed, deployed and tested a lightweight yet effective software for HPE analysis and performance optimization for athletes. MediaPipe's tracking component is optimized for a single person, which implies that even though we obtain these results for single person scenarios, multi-person scenarios may struggle. Despite no elaborate alternatives, we obtained meaningful insights on our sample videos. And, the system still requires access to a basic GPU or NPU, or a decent CPU.

## VI. Mathematical Formulation

This paper presents some mathematical formulas and constraints which are as follows:

*a) Joint Angle Calculation:* Let $A, B, C \in \mathbb{R}^2$represent joint coordinates. The angle at joint $B$is computed as:

$$\theta = \cos^{-1}\left(\frac{(A-B)\cdot(C-B)}{\| A-B \|\| C-B \|}\right)$$

This formulation ensures consistent and rotation-invariant joint angle estimation.

*b) Finite State Machine for Repetition Counting:* Repetition counting is implemented using a finite state machine (FSM) with multiple states depending upon exercise: like 'Up' and 'Down', etc. where transitions are governed by joint angle thresholds like:

- Up → Down when $\theta < \theta_{low}$
- Down → Up when $\theta > \theta_{high}$

A repetition is counted only when a full transition cycle is completed, reducing false positives caused by noise.

## VII. Conclusion And Future Work

This paper presents a survey of the flagship developments in real-time player pose estimation and its limitations, from the traditional marker-based motion capture systems to the modern markerless systems, making data-driven analysis available to a broader spectrum of athletes and the general public. We have also conducted a detailed review of the core technologies, such as the spatially-precise High-Resolution Networks (HRNet) and the temporally aware Transformer models. Furthermore, we highlight the critical optimization strategies, including model compression, lightweight network design for balancing the competing demands of accuracy, latency, and computational costs.

Beyond the survey, this paper also presents a prototype for athletic performance analysis and optimisation. We hope our work will provide useful insights to the readers and motivate further study and innovation in this domain. However, the potential of this prototype is not fully explored with more advanced technologies like MoE, imatrix quantisation and various other frontier models, such as Gemma 4 or Kimi 2.5, which may further improve the performance. Although this demonstrates exciting properties such as scalability, flexibility, and transferability, more research efforts could be made, e.g., exploring the GenAI to demonstrate the insights textually and provide feedback in different complexities based on expert and non-expert users. In addition, we believe our prototype can also be deployed to edge devices where computing power is limited, and power or heat dissipation is constrained.

The future work also includes using wearable Inertial Measurement Units like accelerometers or gyroscopes, and cameras together to correct the other's errors, producing more accurate and robust analysis than either system alone [15]. Future systems could use pose data to build a unique biomechanical model for each athlete, enabling real-time, individualised feedback rather than generic cues. Additionally, integrating AR and VR will elevate perception from 2D screens into immersive virtual space, helping speed up skill acquisition and understanding [17,16,18].